\input phyzzx

\font\mybb=msbm10 at 12pt
\def\bb#1{\hbox{\mybb#1}}
\def\bZ {\bb{Z}}
\def\bR {\bb{R}}


\REF\BST{E. Bergshoeff, E. Sezgin and P.K. Townsend, {\sl Supermembranes and 11
dimensional supergravity}, Phys. Lett. {\bf  189B} (1987) 75.}
\REF\BLT{E. Bergshoeff, L.A.J. London and P.K. Townsend, {\sl Spacetime
scale-invariance and the super p-brane},  Class. Quantum Grav. {\bf 9} (1992)
2545.}
\REF\pol{J. Polchinski and A. Strominger, {\sl New vacua for type-II string
theory}, Phys. Lett. {\bf 388B} (1996) 736.}
\REF\BGPT{E. Bergshoeff, M. de Roo, M.B. Green, G. Papadopoulos and P.K.
Townsend, {\sl Duality of Type II 7-branes and 8-branes}, Nucl. Phys. {\bf B470}
(1996) 113.}
\REF\polwit{J. Polchinski and E. Witten, {\sl Evidence for heterotic-Type I
string duality}, Nucl. Phys. {\bf B460} (1996) 525.}
\REF\strom{A. Strominger, {\sl Open p-branes}, Phys. Lett. {\bf 383B} (1996)
44.}
\REF\pktb{P.K. Townsend, {\sl D-branes from M-branes}, Phys. Lett. {\bf 373B}
(1996) 68.}
\REF\pktc{P.K. Townsend, {\sl Brane surgery}, hep-th/9609217.}
\REF\KM{D. Kutasov and E. Martinec, {\sl M-branes and N=2 Strings},
hep-th/9612102.}
\REF\hewson{S.F. Hewson and M.J. Perry, {\sl The twelve-dimensional super
(2+2)-brane}, hep-th/9612008.}
\REF\dhis{M.J. Duff, P.S. Howe, T. Inami and K.S. Stelle, {\sl
Multimembrane solutions of D=11 supergravity}, Phys. Lett. {\bf 191B} (1987) 
70.}
\REF\pkta{P.K. Townsend, {\sl Worldsheet electromagnetism and the superstring
tension}, Phys. Lett.  {\bf 277B} (1992) 285.}
\REF\azc{J.A. de Azc{\' a}rraga, J.M. Izquierdo and P.K. Townsend, {\sl
Kaluza-Klein origin for the superstring tension},  Phys. Rev. {\bf D45} (1992)
R3321.}
\REF\leigh{R.G. Leigh, {\sl Dirac-Born-Infeld action from dirichlet
$\sigma$-model}, Mod. Phys. Lett. {\bf A4} (1989) 2767.}
\REF\schm{C. Schmidhuber, {\sl D-brane actions}, Nucl. Phys. {\bf B467} (1996)
146.}
\REF\wit{E. Witten, {\sl Bound states of strings and p-branes}, Nucl. Phys. 
{\bf B460} (1995) 335.}
\REF\spec{B. Julia, {\sl Extra-dimensions: recent progress using old ideas}, in
{\it Proceedings of The Second Marcel Grossman Meeting on General Relativity},
pp.79-84 (North Holland 1982); L. Castellani, P. Fr{\'e}, F. Giani, K. Pilch
and P. van Nieuwenhuizen, {\sl Beyond 11-dimensional supergravity, and Cartan
integrable systems}, Phys. Rev. {\bf D26} (1982) 1481; M. Blencowe and M.J.
Duff, {\sl Supermembranes and the signature of spacetime}, Nucl. Phys. {\bf
B310} (1988) 387.}
\REF\revive{C.M. Hull, {\sl String dynamics at strong coupling},
Nucl. Phys. {\bf B468} (1996) 113; C. Vafa, Evidence for F-theory, Nucl. Phys.
{\bf B469} (1996) 403, hep-th/9602022; D. Kutasov and E. Martinec, {\sl New
principles for string/membrane unification}, Nucl. Phys. {\bf B477} (1996) 652;
A. Tseytlin, {\sl Self-duality of Born-Infeld action and Dirichlet 3-brane of
Type IIB superstring theory}, Nucl. Phys. {\bf B469} (1996) 51; I. Bars, {\sl
Supersymmetry, p-brane duality and hidden space-time dimensions}, Phys. Rev.
{\bf D54} (1996) 5203.}
\REF\others{S. Ferrara, R. Minasian and A Sagnotti, {\sl Low-energy
analysis of M and F-theories on Calabi-Yau threefolds}, Nucl. Phys. {\bf B474}
(1996) 323; S. Ferrara, M. Bianchi, G. Pradisi, A. Sagnotti and Y.S. Staney,
{\sl 12-dimensional aspects of 4-dimensional N=1 Type I vacua}, Phys. Lett.
{\bf 387B} (1996) 64.}
\REF\nishino{H. Nishino and E. Sezgin, {\sl Supersymmetric Yang-Mills
Equations in 10+2 Dimensions}, hep-th/9607185.}
\REF\aurilia{A. Aurilia, H. Nicolai and P.K. Townsend, {\sl Hidden 
constants:the
$\theta$-parameter of QCD, and the cosmological constant of N=8 supergravity},
Nucl. Phys. {\bf B176} (1980) 509.}
\REF\schwarz{J.H. Schwarz, {\sl An $Sl(2;\bZ)$ multiplet of Type IIB 
superstrings}, Phys. Lett. {\bf 360B} (1995) 13; {\it erratum, ibid} {\bf
364B}.}
\REF\bergshoeff{E. Bergshoeff, {\sl p-branes, D-branes and M-branes}, 
hep-th/9611099.}


\Pubnum{ \vbox{ \hbox{R/97/22} \hbox{hep-th/9705160} } }
\pubtype{}
\date{}

\titlepage

\title {\bf Membrane tension and manifest IIB S-duality}

\author{P.K. Townsend}
\address{DAMTP, University of Cambridge,
\break
Silver St., Cambridge CB3 9EW, U.K.}

\abstract{A manifestly S-dual, and `12 dimensional', IIB superstring action with
an $Sl(2;\bR)$ doublet of `Born-Infeld' fields is presented. The M-theory
origin of the 12th dimension is the M-2-brane tension, which can be regarded as
the flux of a 3-form worldvolume field strength. The latter is required by the
fact that the M-2-brane can have a boundary on an M-5-brane.}
\endpage

\chapter{Introduction}

In the standard effective action of the 11-dimensional supermembrane [\BST],
i.e. the M-2-brane, the tension is a fixed parameter. In [\BLT] an alternative
action was introduced in which the tension becomes a dynamical variable. This
action is
$$
S= \int\! d^3\xi\, {1\over 2v} \big[ \det g + (\star{\cal G})^2 \big]
\eqn\onea
$$ 
where $\xi$ are the worldvolume coordinates, $v$ is an independent worldvolume
scalar density, $g$ is the induced worldvolume metric, and $\star {\cal G}$ is
the worldvolume dual of the 3-form
$$
{\cal G} = dU - A\, .
\eqn\oneb
$$
The 2-form $U$ is an independent worldvolume gauge potential whereas $A$ is
the pullback of the 3-form gauge potential of D=11 supergravity\foot{We use 
the same letter  to denote superspace forms and their pullbacks since it should
be clear which is intended from the context.}. Thus ${\cal G}$ is a type of 
'modified' field strength 3-form. Note that the action \onea\ is scale
invariant [\BLT], a fact which motivated its construction. Also, for 
an appropriate choice of transformations of $U$, the Lagrangian (and not 
just the action) is invariant under (super)isometries of the background. 
If $g$ and $A$ are interpreted as being induced from superspace tensors 
then the action is $\kappa$-symmetric provided that the background satisfies 
the superfield constraints of D=11 supergravity. 

The $U$ field equation of \onea\ implies that
$$
\star {\cal G}  = Tv
\eqn\onec
$$ 
where $T$ is a constant (with $T\ne0$ spontaneously breaking scale invariance).
The remaining field equations are then equivalent to those of the standard
supermembrane action with tension $T$. Thus, the membrane tension has been
replaced by the  flux of a 2-form gauge potential. This is analogous to the
replacement of the cosmological constant of IIA supergravity by the flux of a
9-form gauge potential [\pol,\BGPT]. In that case, discontinuities in the 
10-form
field strength are associated with domain walls, i.e. 8-branes [\polwit,\BGPT].  One
could similarly associate discontinuities in ${\cal G}$ with boundaries of the
M-2-brane. The fact that the M-2-brane can have a boundary on a 5-brane
[\strom,\pktb] therefore provides a motivation for the new action \onea. Note
that the M-5-brane cannot have a boundary [\pktc] so we should not expect to
have to replace its tension with a 5-form gauge potential. 

The purpose of this article is to point out some consequences, implied by
duality, of the elevation of the M-2-brane tension to the status of a dynamical
variable\foot{Similar considerations were used in [\KM] to motivate a
(2+2)-brane in 10+2 dimensions, also considered in [\hewson] in a similar
context.}. One consequence, following from double dimensional reduction [\dhis],
is that the IIA superstring tension should be replaced by 1-form gauge
potential, as originally advocated in [\pkta] (following an earlier suggestion
[\azc]). It was shown in [\BLT] that this 1-form gauge potential is a worldsheet
Born-Infeld (BI) field. This may sound surprising because BI fields are usually
associated with D-branes rather than `fundamental' strings [\leigh]. In fact,
the quantized flux of the BI field on the D-string can be identified with the
tension of a `fundamental' IIB superstring [\schm]. In other words, the 
D-string
effective action is actually the action for an arbitrary number of `fundamental'
IIB strings bound to a D-string [\wit]. 

Another consequence of the new M-2-brane action \onea\ is that the the
D-2-brane tension must be similarly replaced by a 2-form gauge potential.
T-duality then implies that the tension of each D-p-brane should be replaced by
a p-form worldvolume gauge potential. In particular, the D-string tension should
be replaced by a 1-form gauge field. This cannot be the usual BI field because,
as just noted, its flux is the tension of the `fundamental' string. On the
other hand, IIB S-duality implies that the new 1-form potential must be
exchanged under S-duality with the BI field. In other words, the usual D-string
action should be replaced by a manifestly S-dual one involving an $Sl(2;\bR)$
doublet of 1-form gauge fields. 

As we shall see, the new IIB superstring action is 12-dimensional in an obvious
sense. Supersymmetric theories in twelve dimensions have been the subject of
speculation for a long time [\spec], and the recent suggestions [\revive] of a 
connection with IIB superstring theory have attracted considerable attention 
(see  e.g. [\others,\nishino]). The author sees no direct connection of these 
ideas to the present work, but cannot exclude the possibility. It is perhaps 
worth pointing out that speculations concerning two time directions make sense 
only in the context of a postulated invariance under some 12-dimensional 
orthogonal group, since the number of time directions is related to the 
signature of this group. The new IIB superstring action presented here is only 
$SO(9,1)\times Sl(2;\bR)$ invariant, so the question of whether the twelfth 
dimension is spacelike or timelike does not arise.

\chapter{A manifestly S-dual IIB superstring}

Following the steps in [\pkta,\BLT], it is not difficult to construct a new
manifestly S-dual IIB superstring action. We first introduce two 1-form gauge 
potentials, $V$ and $\tilde V$, and their `modified' 2-form field strength 
2-forms
$$
{\cal F} = dV - B \qquad \tilde {\cal F} = d\tilde V -\tilde B
\eqn\onea
$$
where $V$ and $\tilde V$ are the two 1-form gauge potentials and $B$ and
$\tilde B$ are the pullbacks to the worldvolume of the $NS\otimes NS$ and
$R\otimes R$ two-form gauge potentials, respectively. The worldsheet Hodge
duals 
$\star {\cal F}$ and $\star \tilde {\cal F}$ are worldsheet scalar densities. 
We
can now  write down the manifestly $Sl(2;\bZ)$ invariant (`Einstein frame')
action
$$
S= \int d^2\xi \, {1\over2v}\big\{ \det g + e^{-\phi}[\star{\cal F}]^2 +
e^{\phi}[\star (\tilde {\cal F} - \ell{\cal F})]^2 \big\}\, .
\eqn\onec
$$
The scalar $\phi$ is the IIB dilaton field and $\ell$ the axion. By 
rescaling to the `string-frame' metric one can arrange for ${\cal F}$ to appear
in the BI combination $\det(g+{\cal F})$. Alternatively, one can scale to the
dual D-string frame metric to arrange for $\tilde {\cal F}$ to appear in this
way, so either $V$ or $\tilde V$ may be interpreted as BI fields, but not
both simultaneously. The complex field 
$$
\tau =\ell + i e^{-\phi}
\eqn\oned
$$
transforms under $Sl(2;\bZ)$ via the fractional linear transformation 
$$
\tau \rightarrow {a\tau + b\over c\tau +d}\, , \qquad \pmatrix{a&b\cr c&d}\in
Sl(2;\bZ)\, .
\eqn\onee
$$
The action \onec\ is then invariant if $V$ and $\tilde V$ transform as an
$Sl(2;\bR)$ doublet
$$
\pmatrix{\tilde V \cr V} \rightarrow 
\pmatrix{a&b\cr c&d} \pmatrix{\tilde V \cr V}
\eqn\onef
$$
Since $B$ and $\tilde B$ transform in the same way, the field strength 2-forms
${\cal F}$ and $\tilde {\cal F}$ also form an $Sl(2;\bR)$ doublet. The field
equation of $\tilde V$ implies that
$$
\star (\tilde {\cal F}-\ell {\cal F}) = e^{-\phi} vT
\eqn\oneg
$$
for constant T. If this is substituted into the remaining field
equations\foot{To legitimize substitution into the action one would have to
include a surface term, as discussed in [\aurilia] in a different context; when
this is done one finds the same result as substitution into the field
equations.} one recovers the usual (Einstein frame) super
D-brane equations for a D-string of tension
$T$. 

No attempt will be made here to establish $\kappa$-symmetry. Instead, the
action \onec\ will be interpreted as a purely bosonic one. Passing to the
Hamiltonian form we then find the equivalent (bosonic) action
$$
S= \int dt\oint d\sigma \big\{ \dot x\cdot p + (\partial_t V_1) E +
(\partial_t \tilde V_1) \tilde E +  V_0 E' + \tilde V_0 \tilde E' + 
s\, x'\cdot p -
{1\over2} u\, {\cal H}\big\}
\eqn\onega
$$
where $s$ is a Lagrange multiplier (shift function), $u= v/(x')^2$ is another
Lagrange multipler (lapse function), and $E$ and $\tilde E$ are the electric 
field variables conjugate to $V_1$ and $\tilde V_1$, respectively. The
Hamiltonian constraint function ${\cal H}$ is
$$
{\cal H} = \big(p+ \tilde E \tilde {\cal B} + E {\cal B}\big)^2  + (x')^2 [
e^\phi (E+\ell\tilde E)^2 + e^{-\phi}\tilde E^2 ] 
\eqn\oneh
$$
where
$$
{\cal B}_\mu = (x')^\nu B_{\mu\nu} \qquad 
\tilde {\cal B}_\mu = (x')^\nu \tilde B_{\mu\nu}\, .
\eqn\onei
$$
A prime indicates differentiation with respect to the string's spatial
coordinate $\sigma$. The constraints imposed by $V_0$ and $\tilde V_0$ imply 
that the electric fields $E$ and $\tilde E$ are independent of $\sigma$.
Variation with respect to $V_1$ and $\tilde V_1$ shows that $E$ and $\tilde E$
will remain at their initial values. If $V$ and $\tilde V$ are taken to be
$U(1)$ gauge fields then the values allowed to $E$ and $\tilde E$ are quantized.
We shall suppose that the units are such that $E$ and $\tilde E$ are 
integers:
$$
E= m\, , \qquad \tilde E= n\, .
\eqn\onej
$$
If we now use this in \onega, and discard surface terms\foot{If the action is
invariant under some symmetry, e.g. supersymmetry, then the process of
discarding surface terms may lead to an action that is invariant up to a
surface term. This is why the supermembrane action in flat space, for which
the supersymmetry variation of the Wess-Zumino term is a surface term, can be
replaced by the action \onea\ for which the Lagrangian, and not just the action, is invariant.}, we arrive at the action
$$
\eqalign{
S= &\int dt\oint d\sigma \big[ \dot x\cdot p +  s\, x'\cdot p  \cr
&\qquad -{1\over2} u
\{ \big(p+ m\tilde {\cal B} + n{\cal B}\big)^2  + (x')^2 [
e^\phi (m+n\ell)^2 + e^{-\phi}n^2 ]\}\big]\, .}
\eqn\onek
$$
This is the hamiltonian form of the action for an $(n,m)$ string. Setting
$e^\phi = g_s$ (the string coupling) we find that the tension {\it in the 
string frame} is 
$$
T= \sqrt{ (n/g_s)^2 + (m+ n\ell)^2}
\eqn\onel
$$
as expected [\schwarz].

Note that the action \onega\ takes the form
$$
S=\int dt\oint d\sigma \, \big[ \dot X\cdot P -\lambda^I {\cal H}_I\big]
\eqn\onem
$$
where $X= (x^\mu, V_1,\tilde V_1)$ and $P=(p_\mu, E,\tilde E)$ and ${\cal H}$
are a set of (first class) constraints. Thus, the action is 12 dimensional in
an obvious sense. We conclude that {\it the M-theory origin of the 12th
dimension of IIB superstring theory is the M-2-brane tension}. 

\chapter{p-brane boundaries and worldvolume p-forms}
 
It was noted above that, given a 2-form potential on the worldvolume of
the D-2-brane, T-duality requires the existence of a $p$-form gauge potential on
the worldvolume of each D-p-brane. The flux of its $(p+1)$-form field strength
through the worldvolume equals the D-p-brane tension. We have already used the
fact that IIB S-duality requires a 1-form potential ($\tilde V$) on the IIB
superstring worldsheet; similar reasoning shows that the IIB $NS\otimes NS$, 
or `solitonic' 5-brane must have a 5-form gauge potential. In fact, once we
introduce the 2-form gauge potential for the M-2-brane, the existence of a
p-form gauge potential on almost all other p-branes follows by duality. An
exception is the M-5-brane. Given a 4-form potential on the 4-brane worldvolume
we {\it cannot} deduce the existence of a 5-form potential on the M-5-brane
worldvolume because the former has an alternative 11-dimensional explanation.
This is just as well since we argued earlier that the M-5-brane action should
not have such a field. 

To see how the absence of a 5-form gauge field on the M-5-brane is compatible
with the occurrence of a 4-form potential on the D-4-brane obtained by
double-dimensional reduction, we note that $x^{11}$ may first be replaced by
its  4-form dual with 5-form field strength. The double-dimensional reduction
ansatz now corresponds to a non-vanishing flux of this 5-form field strength
through the D-4-brane worldvolume, so we may identify the 4-form potential on
the D-4-brane as the dual of the M-5-brane field $x^{11}$. Similarly, a 2-form
potential on the D-membrane is not implied by a 1-form on the IIA superstring
(although the reverse implication is valid) but it {\it is} implied by the
existence of a 2-form on the D-2-brane, and the latter is implied by a
combination of T-duality and IIB S-duality given the BI field on the D-string.
Thus, by reversing the previous logic, we can use duality to deduce the 
existence
of the 2-form gauge potential on the M-2-brane from known results on D-branes,
but we cannot similarly deduce the existence of a 5-form gauge potential on the
M-5-brane. For example, while the latter would be implied by a 5-form potential
on the
$NS\otimes NS$, or `solitonic', 5-brane of the IIA theory there is no reason 
(in
contrast to the IIB case) to suppose that there is such a field. Once one
accepts the hypothesis that the M-2-brane has a 2-form gauge potential but the
M-5-brane does not have a 5-form gauge potential it follows by duality\foot{At
least for $p\le6$. Formally one could use T-duality to conclude that a D7-brane can have a boundary on a D9-brane but it is not clear to the author how this 
should be interpreted.} that a {\it a p-brane has a p-form gauge potential if 
and only if it can have a boundary on some other brane}.

Finally, we wish to point out that the results reported here will likely have
implications for one of the outstanding unsolved problems in the ongoing
program to determine the full $\kappa$-symmetric actions of all superstring and
M-theory p-branes, namely the IIB solitonic 5-brane. As we have seen, this
action should have a 5-form gauge potential. It is tempting to suppose that
there is a manifestly S-dual IIB 5-brane action analogous to the IIB string
action given here but with an $Sl(2;\bR)$ doublet of 5-form gauge potentials.
Note that there cannot be an $Sl(2;\bR)$ doublet of BI gauge fields in this
case because a second BI field would disturb the balance of degrees of freedom.
One suspects, therefore, that this manifestly S-dual IIB 5-brane action must
involve the one BI field and its 3-form dual [\bergshoeff] in a symmetric way. 
If so this
could make the action difficult to find. 

\vskip 1cm
{\bf Acknowledgements}. I thank G. Papadopoulos for helpful
conversations.

\refout

\end